\documentclass[11pt,showpacs,preprint]{revtex4}
\usepackage{graphicx}
\usepackage{dcolumn}
\usepackage{bm}
\begin{document}

\title{Relativistic effects in exclusive neutron-deuteron breakup}

\author{R.~Skibi\'nski}
\affiliation{M. Smoluchowski Institute of Physics, Jagiellonian
University,
                    PL-30059 Krak\'ow, Poland}

\author{H.~Wita{\l}a}
\affiliation{M. Smoluchowski Institute of Physics, Jagiellonian
University,
                    PL-30059 Krak\'ow, Poland}

\author{J.~Golak}
\affiliation{M. Smoluchowski Institute of Physics, Jagiellonian
University,
                    PL-30059 Krak\'ow, Poland}

\date{\today}

\begin{abstract}
 We  extended the study of relativistic effects in neutron-deuteron 
scattering to the exclusive breakup. To this aim we 
solved the  three-nucleon Faddeev
 equation including  such relativistic features as relativistic kinematics 
 and boost effects 
 at  incoming neutron lab. energies  $E_n^{lab}=65$~MeV, $156$~MeV and 
$200$~MeV.
 As dynamical
 input a relativistic nucleon-nucleon interaction exactly on-shell
 equivalent to the CD~Bonn potential has been used. 
 We found that the magnitude of 
relativistic effects increases with the incoming neutron energy and, depending 
on the phase-space region, relativity can increase as well as decrease the 
nonrelativistic breakup cross section. 
 In some regions of the breakup phase-space dynamical boost effects are important. 
 For a number of measured exclusive cross sections 
 relativity seems to improve 
 the description of data.
\end{abstract}


\maketitle
\setcounter{page}{1}

\section{Introduction}

The high precision nucleon-nucleon (NN) potentials which describe  
very well the NN data set up to about $350$~MeV~\cite{AV18,CDBONN,NIJMI} 
 form  a very firm basis for 
a study of three-nucleon (3N) reactions. Powerful computers and development  
of modern algorithms provided numerically exact solutions of the 3N 
Schr\"odinger equation both in the momentum and coordinate representation. 
This permitted theoretical calculations for cross sections and spin 
observables in elastic nucleon-deuteron (Nd) scattering and breakup 
processes with different dynamical assumptions about underlying nuclear 
forces~\cite{glo96}. 
With increasing amount of precise 3N elastic scattering data it 
turned out, that nonrelativistic description based on pairwise forces only is 
insufficient to explain the data at higher energies of the 3N 
system. 
Adding a 3N force (3NF) to the pairwise interactions led in many cases 
to a better description of the 
data~\cite{wit98,sek02,wit01,abf98,wit99,hat02,cad01,bieber2000,erm2001,erm2003,erm2005}.  
However, at energies higher than $\approx 100$~MeV current 3NFs only 
partially improved the description of data, leaving in some cases   
discrepancies which  were comparable in magnitude 
to the effects of the 3NFs 
themselves~\cite{sek02,wit01,abf98,wit99,hat02,cad01,bieber2000,erm2001,erm2003,erm2005}. 

This situation triggered  investigations of the 3N continuum taking 
relativistic effects into account. In ref.~\cite{witrel} we 
performed such a study 
for Nd elastic scattering. We extended the Hamiltonian scheme in equal 
time formulation worked out in~\cite{kam2002} to 
the 3N scattering taking as a starting point the 
Lorentz boosted NN potential which generates the NN t-matrix in a moving frame 
via a standard Lippmann-Schwinger equation. The NN 
potential in an arbitrary moving frame is based on the interaction in the 
two-nucleon c.m. system, which enters a relativistic NN Schr\"odinger or 
Lippmann-Schwinger equation. The relativistic equation differs from the 
nonrelativistic one just by the relativistic form of the kinetic energy. 
We constructed the relativistic two nucleon (2N) potential by performing  
an analytical scale transformation of momenta, which relates NN potentials 
in the nonrelativistic and relativistic Schr\"odinger equations in such a way, 
that exactly the same NN phase shifts are obtained by both 
equations~\cite{kam98}. 

In our first study~\cite{witrel} we looked for changes 
in elastic neutron-deuteron (nd) 
scattering observables when the nonrelativistic form of the kinetic 
energy was replaced by the relativistic one and a proper treatment of 
boost effects and Wigner rotations of spin states was included. 
It turned out, that the effects of spin rotations in the studied 
energy range up to $250$~MeV were practically negligible for elastic 
scattering cross sections and analyzing powers. The relativistic effects 
for the elastic scattering 
 cross section were significant only at higher energies and  
restricted to the very backward angles, where relativity increased 
the nonrelativistic cross section. The decisive role was played by the boost 
effects which reduced the transition matrix elements at higher energies and 
led, in spite of the increased elastic scattering 
 relativistic phase-space factor as compared to the nonrelativistic one, to 
rather small effects in the cross section. 

In view of upcoming measurements of the higher energy Nd breakup process we would 
like to extend the study of ref.~\cite{witrel} to this reaction. 
The first results presented in~\cite{wit2006} revealed a unique 
selectivity of the 
complete breakup reaction useful to investigate the pattern of relativistic 
effects and to study their significance. 
Here we would like to study how 
relativistic effects are distributed over the breakup phase-space and if and to  
what extent the existing data verify the predicted effects.

The paper is organized as follows. 
In Sec. II for convenience of the reader we shortly 
explain  the relativistic features
underlying our treatment of a relativized Faddeev 
equation in the 3N continuum. The very detailed presentation which 
incorporated the definition of the boosted two-body force, 
the various two-and three-body
states in general frames, the Wigner rotations and the 
singularity structure of the
relativistic free 3N propagator were given in \cite{witrel}. 
In Sec. III we give relativistic and 
nonrelativistic formulas for the transition matrix elements and the 
breakup cross section.  
In Sec. IV we apply our
formulation using  a relativistic NN interaction which is exactly
 on-shell
equivalent to the nonrelativistic CD~Bonn potential and
 solve the relativized
3N Faddeev equation with  different  approximations for the boost.
 At two incoming  neutron energies, 65 and 200 MeV, 
we look for the magnitude and the distribution pattern of 
relativistic effects over the entire phase-space of the breakup reaction. 
For the cases where the data are available, we compare them to theoretical predictions. 
Finally, Sec. V contains a summary and outlook.

\section{Formulation}

When nucleons interact through a NN potential $V$, the breakup operator T
satisfies the Faddeev-type integral equation~\cite{wit88,glo96}
\begin{eqnarray}
T\vert \phi >  &=& t P \vert \phi > + t P G_0 T \vert \phi > .
\label{eq1a}
\end{eqnarray}
The 2N t-matrix t results from the interaction $V$ through
the Lippmann-Schwinger equation and the permutation operator  
 $P=P_{12}P_{23} + P_{13}P_{23}$ is given in terms
of a transposition $P_{ij}$ which interchanges nucleons i and j. 
The incoming state  
$ \vert \phi > =
\vert \vec q_0 > \vert \phi_d > $ describes the free
nucleon-deuteron motion with relative momentum $\vec q_0$ and
the deuteron wave function $\vert \phi_d >$. Finally $G_0$ is the free 
3N propagator. 

The full nd breakup  transition operator $U_0$    
 is  given in terms of T by~\cite{wit88,glo96}
\begin{eqnarray}
U_0  &=& (1 + P) T  .
\label{eq1cc}
\end{eqnarray}

This is our standard nonrelativistic formulation, which is equivalent to the
nonrelativistic 3N Schr\"odinger equation and respects the boundary conditions.
The formal  structure of these equations  in the relativistic case remains
the same but the ingredients change. As explained in ~\cite{relform} 
the relativistic 3N Hamiltonian has the same form as the nonrelativistic one,
 only the momentum dependence of the kinetic energy changes and the relation
of the pair interactions to the ones in their corresponding c.m. 
frames changes,
too. Consequently all the formal steps 
leading to Eqs.(\ref{eq1a}) and (\ref{eq1cc}) remain the same.

The relativistic kinetic energy of three equal 
mass (m) nucleons  in their c.m. system can be written in terms of 
 the momentum dependent 2N mass operator 
 $2\omega({\vec k}) \equiv 2\sqrt{m^2 + {\vec k}^{\ 2}}$ 
 and  the momentum of the third nucleon  $\vec q$  as 
\begin{eqnarray}
H_0 &=& \sqrt{(2\omega({\vec k}))^2 + {\vec q}^{\ 2}} 
+ \sqrt{m^2 + {\vec q}^{\ 2}} .
\label{eq1an1}
\end{eqnarray}
Here  $\vec k$ and $-\vec k$ are  
the momenta of two nucleons in one of the two-body c.m. subsystems 
 and $-\vec q$ is the total momentum of
this chosen two-body subsystem. Any of the  
three possible two-body subsystems 
can be taken. 

The boosted 2N potential  in
the 2N frame moving with momentum $\vec q$ is taken as 
~\cite{relform1} 
\begin{eqnarray}
V(\vec q \, ) \equiv \sqrt{( 2\omega(\vec k) + v)^2 + {\vec q}^{\ 2}} 
 - \sqrt{ (2\omega(\vec k))^2 + {\vec q}^{\ 2}} , 
\label{eq1}
\end{eqnarray}
where $ V(\vec q)$ 
for $\vec q = 0$
reduces to the potential $v$ defined in the 2N 
  c.m. system.  Note that also in that system the relativistic kinetic energy
of the two nucleons has to be chosen, which together with $v$ defines the 
interacting two-nucleon mass operator occurring in Eq.(\ref{eq1}).

The boosted 2N t-matrix $t(\vec k, \vec k~'; \vec q~)$  fulfills the 
 relativistic
2N Lippmann Schwinger equation, which
in a general frame reads
\begin{eqnarray}
 t( \vec k, \vec k~' ;  \vec q~) &=& 
 V( \vec k, \vec k~' ;  \vec q~) + 
\int d^3k'' { 
{V( \vec k, \vec k~'' ;  \vec q~)
 t( \vec k~'', \vec k~' ;  \vec q~) } 
\over{ \sqrt{ {( 2\omega( {\vec k}^{\ '} )^{\ 2} + {\vec q}^{\ 2}} }  
-  \sqrt{ {(2\omega(  {\vec k}^{\ ''})^{\ 2} 
+ {\vec q}^{\ 2}} } + i\epsilon  } } .
\label{eq2a}
\end{eqnarray}

 Using Eq.~(\ref{eq1}) the relativistic 
 2N Schr\"odinger equation for the deuteron
 in a moving frame 
 can be cast into the form
\begin{eqnarray}
 \phi_d( \vec k) &=& 
{ {1} \over { \sqrt{ {M_d^2 + {\vec q}^{\ 2}_0}} -  
\sqrt{ {( 2\omega (\vec k))^2 + {\vec q}^{\ 2}_0} } } } 
\int d^3k' 
 V( \vec k, \vec k~' ;  \vec q_0 ) 
 \phi_d( \vec k~') ,  
\label{eq2b}
\end{eqnarray}
where $M_d$ is the deuteron rest mass and 
$\sqrt{ {M_d^2 + {\vec q}^{\ 2}_0}}$ its energy in the moving frame.  

The new relativistic ingredients in Eq.(\ref{eq1a})  will therefore  
be the boosted
 t-operator and the relativistic 3N propagator
\begin{eqnarray} 
G_0 &=& {{1}\over{E + i\epsilon - H_0}} ,
\label{eq2c}
\end{eqnarray}
where $H_0$ is given in Eq.~(3) and $E$ is the total 3N c.m. 
energy expressed in terms of the initial
neutron momentum $\vec q_0$ relative to the deuteron
\begin{eqnarray}
E &=& \sqrt{(M_d)^2 + {\vec q}_0^{\ 2} } + \sqrt{m^2 + {\vec q}_0^{\ 2}} .
\label{eq2d}
\end{eqnarray}

We solve numerically Eq.(\ref{eq1a}) in its nonrelativistic or relativistic 
form for
any NN interaction using a momentum  space partial wave decomposition. 
In the nonrelativistic case the partial wave projected momentum space basis is
 taken as $\vert p q \alpha > \equiv  \vert p q (ls)j 
(\lambda{1\over{2}})IJ(t{1\over{2}})T >$, 
 with the magnitudes p and q of standard Jacobi momenta (see \cite{book}) 
and $(ls)j$ two-body quantum numbers with obvious meaning. The quantum 
numbers 
$(\lambda 1/2)I$ refer to the third nucleon (its motion is described by the momentum q), and 
$J$ is the total 3N angular momentum. The subsystem isospin $t$ couples 
with the spectator isospin to the total 3N isospin $T$. 

In the relativistic case the nonrelativistic relative 
two-nucleon momentum
$\vec p$ is replaced by $\vec k$, where 2N c.m. momenta $\vec k$ and 
$-\vec k$ are related to general momenta of these two
nucleons, say $\vec p_2$ and $\vec p_3$,  
by a Lorentz boost:
\begin{eqnarray}
\vec k &\equiv& \vec k(\vec p_2, 
\vec p_3) \cr
&=& \frac{1}{2} (\vec p_2 - \vec p_3 - 
\vec p_{23} \frac{E_2 - E_3}{E_2 + E_3 + \sqrt{(E_2 + E_3)^2 - 
\vec p_{23}^{~2}}  } ) ,
\label{eq_n2}
\end{eqnarray}
with $E_i = \sqrt{m^2 + {\vec p}_i^{\ 2}}$ and
 $\vec p_{23} = \vec p_2 + \vec p_3$. 
The third spectator nucleon has 
momentum $\vec p_1$ which together with 
the total two-nucleon momentum $\vec p_{23}$ adds up  
to zero in the 3N c.m. system. It is the momentum $\vec q = \vec p_1$ 
which replaces the nonrelativistic Jacobi momentum $\vec q$ in the 
relativistic case to describe unambiguously a 3N configuration. 

The construction of the momentum space partial wave basis in the relativistic 
case starts with definition of the 2N subsystem partial wave state 
 $\vert (ls)jk \mu; \vec 0>$ defined in the 2N c.m. subsystem~\cite{witrel}. 
Boosting this state to the 3N c.m. system along the total momentum  $\vec p_{23}$ 
and coupling it there with the state of the spectator nucleon leads to 
a 3N partial wave state 
$\vert k \;q=p_1\; \alpha > \equiv 
\vert k p_1  (ls)j (\lambda {1\over{2}})I (jI) JM > 
\vert (t{1\over{2}})TM_T>$.  Since the performed boost generally is not 
parallel  to the momenta $\vec k$ and $-\vec k$ of the two nucleons in their 
2N c.m. system it leads to Wigner rotations of their spin states. This 
rotation complicates the evaluation of the partial wave 
representation of the permutation operator P in the 
basis $\vert k\; q=p_1\; \alpha >$.
 In \cite{witrel} it was shown that this representation can be given 
 in a form
 which resembles closely the
one appearing in the nonrelativistic case ~\cite{book,glo96} 
\begin{eqnarray}
< k ~ q ~ \alpha \vert ~ P ~ \vert k' ~ q' ~ \alpha' > &=& 
\int_{-1}^{1} dx { {\delta(k-\pi_1)} \over { k^{l+2} }  } ~ 
{ {\delta(k'-\pi_2)} \over { k'^{l'+2} }  } ~  \cr 
&~& 
{1\over {N_1(q, q',x) } } ~ {1\over {N_2(q, q',x) } } ~ 
G_{\alpha \alpha'} (q, q', x) ,
\label{eqm7}
\end{eqnarray}
what leads after projecting Eq.~(\ref{eq1a}) onto $\vert k q \alpha >$ to 
 an infinite system of coupled integral equations analogous to
 the nonrelativistic one~\cite{wit88,glo96}:
\begin{eqnarray}
< k q \alpha \vert T(E) \vert  \phi >  &=& < k q \alpha \vert t P \vert \phi >
 + \sum_{\alpha ~'} \sum_{l_{\bar {\alpha}} } \int_0^{\infty} dq' q'^2 
\int_{-1}^{1} dx  {{<kl_{\alpha} \vert t^{(\alpha)}(E 
- \sqrt{m^2 + q^2}) \vert \pi_1 l_{\bar {\alpha}}  >}
\over
{\pi_1^{l_{\bar {\alpha}}}  }} \cr
&~& \times { 
{ G_{\bar {\alpha} \alpha' }(q, q', x) }
\over {   {N_1(q, q',x) }  ~  {N_2(q, q',x) }  ~  }} 
{ {  < \pi_2 q' \alpha' \vert T(E) \vert  \phi >  }
\over{  {\pi_2}^{l_{\alpha'}} }} \cr
&~& \times {{A }\over{  x_0 + i\epsilon - x }} .
\label{eq1f}
\end{eqnarray}

The geometrical coefficients  $G_{\bar {\alpha} \alpha' }(q, q', x)$, 
the coefficients  $N_1(q, q',x)$ and  $N_2(q, q',x)$,  and
the momenta $\pi_1$ and $\pi_2$ stem
from the matrix element  $<kq \alpha \vert P \vert k'q'\alpha'>$ of the
permutation operator (Eqs. (C6-C8) in ref~\cite{witrel}). 
 The last part ${{A }\over{  x_0 + i\epsilon - x }}$ results from the 
free propagator (Eqs.(34-35) in ref.~\cite{witrel}). 
The quantum numbers in the set $\bar {\alpha}$ differ from those in $\alpha$ 
only in the orbital angular momentum $l$ of the pair. 

The main difficulty in  treating Eq.(\ref{eq1f}) is caused by
the singularities of the free propagator $G_0$ which occur in the region
of q and q' values for which $\vert x_0 \vert \le 1$. In addition, at $q
= q_0$ there
is the singularity of the 2N t-matrix  in the $^3S_1-^3D_1$ partial wave
state, where the deuteron bound state exists. 
 How to treat those  singularities is described in detail in
\cite{witrel,wit88}. 
 Equation (\ref{eq1f}) is solved by generating its Neumann
series, which is then summed up by the Pad\'e method~\cite{wit88}. 

Due to its short-range nature, the NN force can be considered
negligible beyond a certain value $j_{max}$ of the total angular momentum
in the two nucleon subsystem. Generally with increasing energy
$j_{max}$ will also increase. For $j > j_{max}$ we put the t-matrix to be zero,
which yields a finite number of coupled channels for each total angular
momentum J and total parity $\pi=(-)^{l+\lambda}$ of the 3N system. 
To achieve converged results at our energies we used all partial 
wave states with total
angular momenta of the 2N subsystem  up to $j_{max}=5$ and took into account 
 all total
angular momenta of the 3N system up to $J=25/2$. This leads to a system
of up to 143 coupled integral equations in two continuous variables
for a given $J$ and parity.

As  dynamical input we used a relativistic interaction $v$, 
which is defined as partner of the relativistic kinetic energy, 
generated from 
the nonrelativistic NN potential CD~Bonn 
according to the analytical prescription of ref.~\cite{kam98}. 
 In ref.~\cite{kam2002}  it 
was shown that the explicit calculation of the matrix elements 
$V( \vec k, \vec k~' ;  \vec q~ )$
according to Eq.~(4) for the boosted potential
requires the knowledge of the NN bound state wave function and the
half-shell NN t-matrices in the 2N c.m. system. 
Here, as in our study of the elastic nd scattering~\cite{witrel},  
we do not treat the boosted potential matrix element 
 in all its complexity as given in ref.~\cite{kam2002} but 
restrict ourselves to the leading order term in a $q/\omega$ and 
$v/\omega$ expansion  
\begin{equation}
V(\vec k, \vec k~'; \vec q~)  = v(\vec k, \vec k') \times [~ 1  
 -  \frac{{\vec q}^{\ 2}} {8\sqrt{m^2 + {\vec k}^{\ 2}}  
\sqrt{m^2 +  ( {\vec k}^{\, \prime} )^2 }} ~ ] .
\label{eq2apr}
\end{equation}

In order to study importance of the boost effect we will present in 
 addition to this approximation  also  the
results for two more drastic approximations. 
In the first one  the boost effects 
are neglected completely  
\begin{eqnarray}
V(\vec k, \vec k~'; \vec q~)  &=&  
v(\vec k, \vec k~') ,
\label{ap1}
\end{eqnarray}
and in the second one the k-dependence of the first order 
relativistic correction term is 
omitted 
\begin{eqnarray}
V(\vec k, \vec k~'; \vec q~)  &=&  
v(\vec k, \vec k~')~ \left(~ 1 - 
\frac{{\vec q}^{\ 2}}{8m^2 } ~ \right) .
\label{ap2}
\end{eqnarray} 

The quality of these approximations can be checked 
by calculating
the deuteron wave function $\phi_d(\vec k)$ for the  deuteron 
moving with  momentum $\vec q$ using Eq.(\ref{eq2b}). This 
wave function  depends only on the 2N c.m. relative 
momentum $\vec k$ inside the deuteron 
and is thus independent of  the boost momentum 
$\vec q$. 
When  the boost effects are fully taken into account the 
solution of Eq.(\ref{eq2b}) must provide exactly the 
deuteron binding energy and the D-state probability equal to the values
for the deuteron at rest. 
 We checked in \cite{witrel}, 
that  neglecting the boost totally or 
omitting the k-dependence of the first order term 
 are poor approximations, especially at the higher energies ($250$~MeV) we 
studied. 
In contrast, the approximation given in Eq.(\ref{eq2apr}) appears acceptable,
even for the strongest boosts, reproducing closely the deuteron binding 
energy and the D-state probability for the deuteron at rest. 
Relying on that result we have chosen the expression (\ref{eq2apr}) 
for the boosted potential in the following investigations.
 Since the solution of the 3N relativized Faddeev 
 equation including Wigner spin 
rotations due to complicated form of $G_{\alpha \alpha' }$ in Eq.(\ref{eqm7})
requires much more computer time and since 
 we restrict in this 
study to the breakup unpolarized cross sections only we neglected the Wigner rotations 
completely.

\section{Relativistic and nonrelativistic breakup cross section}

The exclusive breakup measurements $d(n,N_1~N_2)N_3$ are performed in the lab. 
system with two of the three outgoing nucleons ($N_1$ and $N_2$) 
detected in coincidence by 
detectors placed at fixed angles ($\theta_1^{lab},\phi_1^{lab}$) and 
($\theta_2^{lab},\phi_2^{lab}$) and their kinetic energies $E_1^{lab}$ and 
 $E_2^{lab}$  measured. 
The experimental events are then located in the $E_1^{lab} - E_2^{lab}$ 
energy plane along a kinematical curve determined by the energy and momentum 
conservation. For each point on this curve the nucleons have  definitive 
 momenta. The breakup observables are normally shown as a function of 
an arc-length S of that kinematical locus 
 (the starting point of which is defined according 
to some convention) or as a function of energy $E_1^{lab}$. Theoretical 
predictions  for different observables are obtained from  the matrix elements 
of the breakup transition operator (Eq.\ref{eq1cc})
\begin{eqnarray}
< \phi_0 \vert U_0 \vert \phi >  &=&  
< \phi_0 \vert (~1 + P~)T \vert \phi > ,
\label{eobs1}
\end{eqnarray} 
where the  state 
 $\vert \phi_0 >=\vert \vec p_1 \vec q_1 m_1 m_2 m_3 \nu_1 \nu_2 \nu_3 >$ 
 describes the relative motion of free outgoing nucleons 
and specifies their spin ($m_i$) and isospin ($\nu_i$) magnetic 
quantum numbers. 
 This relative motion is described by standard Jacobi momenta 
($\vec p_i, \vec q_i$)~\cite{glo96,book} which are given 
 in terms of  individual 
momenta $\vec k_i$ of the three nucleons in a particular kinematically 
complete breakup configuration by $i,j,k=1,2,3$ and cyclic permutations
\begin{eqnarray}
\vec p_i  &=& {1 \over {2}} (\vec k_j - \vec k_k)  \cr 
\vec q_i &=& {2 \over {3}} (\vec k_i - {1\over {2}} (\vec k_j + \vec k_k)) .
\label{eobs3}
\end{eqnarray}

Applying in Eq.(\ref{eobs1}) the permutation operator to the left provides 
 three contributions to the breakup amplitude
\begin{eqnarray}
_1< \phi_0 \vert (1+P)T \vert \phi >  &=& 
 _1< \vec p_1 \vec q_1 m_i \nu_i \vert T \vert \phi > + 
 _2< \vec p_1 \vec q_1 m_i \nu_i \vert T \vert \phi > + 
 _3< \vec p_1 \vec q_1 m_i \nu_i \vert T \vert \phi > \cr 
 &=&  _1< \vec p_1 \vec q_1 m_1 m_2 m_3 \nu_1 \nu_2 \nu_3 \vert T \vert \phi > \cr  
&+& 
 _1< -{1\over 2}\vec p_1 - {3\over 4} \vec q_1, 
 \vec p_1 -{1\over 2} \vec q_1,  m_2 m_3 m_1 \nu_2 \nu_3 \nu_1 \vert T \vert \phi > \cr 
&+& 
 _1< -{1\over 2}\vec p_1 + {3\over 4} \vec q_1, 
 -\vec p_1 -{1\over 2} \vec q_1,  m_3 m_1 m_2 \nu_3 \nu_1 \nu_2 \vert T \vert \phi > ,
\label{eobs4}
\end{eqnarray}
where the subscripts on the left side of the matrix elements indicate the leading nucleon. 

For the nonrelativistic case  invariance of Jacobi momenta and amplitudes 
$< p q \alpha \vert T \vert \phi >$ permits calculating observables 
directly in the 3N lab. system. 
 However, in the relativistic case the  amplitudes 
 $< k q \alpha \vert T \vert \phi >$ 
are provided  in the 3N c.m. system and 
the transition matrix elements are first 
calculated in that system. In order to compare with the 
data measured in the lab. system, the proper Lorentz transformation from 
 the 3N c.m. to the lab. system must be performed. 

In the 3N c.m. system each term in Eq.(\ref{eobs4}) is given by
\begin{eqnarray}
_1< \vec p_1 \vec q_1 \vert T \vert \phi >  &=& 
 \frac{1}{ N(\vec k_2, \vec k_3) } < \vec k(\vec k_2, \vec k_3), \vec k_1 
 \vert T \vert \phi >, 
\label{eobs5}
\end{eqnarray}
where 
\begin{eqnarray}
 N(\vec k_2, \vec k_3)  &=&
\sqrt{ \frac{4 E_2 E_3} {  (E_2 + E_3) \sqrt{(E_2 + E_3)^2 - 
 (\vec k_2 + \vec k_3)^2   }  }     } .
\label{eobs6}
\end{eqnarray}

Assuming that the incoming beam of neutrons with the spin projection $\mu$ 
 moves along the z-axis, 
 the matrix element $< \vec k(\vec k_2, \vec k_3), \vec k_1 
 \vert T \vert \phi >$ follows from the calculated amplitudes 
 $< k q \alpha \vert T \vert \phi >$ by
\begin{eqnarray}
&& < \vec k(\vec k_2, \vec k_3), \vec k_1 
 \vert T \vert \phi >  =
 \sum_{J,\pi} \sum_{l\lambda L} 
Y_{l \lambda}^{L \mu+m_d-m_1-m_2-m_3}(\hat k, \hat k_1) 
\sum_{jsItS} \sqrt{\hat j \hat I \hat L \hat S} 
\left \lbrace \matrix{ l & s & j \cr
                       \lambda & {1\over {2}} & I \cr
                       L & S & J \cr} \right \rbrace  \cr \nonumber 
&& (LSJ \vert \mu+m_d-m_1-m_2-m_3,m_1+m_2+m_3,\mu+m_d) 
(s {1 \over 2} S \vert  m_2+m_3,m_1,m_1+m_2+m_3) \cr \nonumber 
&& (t {1\over 2} T \vert \nu_2+\nu_3,-{1\over 2}) 
({1\over 2} {1\over 2} t \vert \nu_2,\nu_3, \nu_2+\nu_3) 
 <k~ k_1 ~\alpha \vert T \vert \phi >,
\label{eobs7}
\end{eqnarray}
where $m_d$ is the spin projection of the deuteron.

The 3N c.m. breakup amplitude $< \phi_0 \vert (1+P)T \vert \phi >$ 
 provides unpolarized cross section 
in this system. Thereby the exclusive relativistic 
cross sections along S-curve 
 $({d^5\sigma \over 
{d\Omega_1d\Omega_2dS}})^{(c.m.)}$ or projected on the $E_1$ axis 
 $({d^5\sigma \over {d\Omega_1d\Omega_2dE_1}})^{(c.m.)}$ are given by
\begin{eqnarray}
&& ({d^5\sigma \over {d\Omega_1d\Omega_2dS}})^{(c.m.)} = (2\pi)^4{1\over{2 \cdot 3}} 
\sum_{m_i m_f} \vert < \phi_0 \vert (1+P)T \vert \phi > {\vert}^2 
{1\over I} {m_1^{in} \over {E_1^{in}}} {m_2^{in} \over {E_2^{in}}}
{m \over {E_3} } { {m^2p_1p_2} \over { \sqrt{({\tilde A}^2 
+ {\tilde B}^2)} } } ,
\label{eobs8}
\end{eqnarray}
and
\begin{eqnarray}
&& ({d^5\sigma \over {d\Omega_1d\Omega_2dE_1}})^{(c.m.)} = (2\pi)^4{1\over{2 \cdot 3}} 
\sum_{m_i m_f} \vert < \phi_0 \vert (1+P)T \vert \phi > {\vert}^2 
{1\over I} {m_1^{in} \over {E_1^{in}}} {m_2^{in} \over {E_2^{in}}}
{m \over {E_3} } { {m^2p_1p_2} 
\over { \vert {\tilde B} \vert } } ,
\label{eobs9}
\end{eqnarray}
with the (invariant) flux I 
\begin{eqnarray}
I &\equiv& \vert {\vec p_1^{~in} \over {E_1^{in}}} - 
{\vec p_2^{~in} \over {E_2^{in}}}\vert 
 =  { 1 \over {E_1^{in} E_2^{in} }} \sqrt{(E_1^{in} E_2^{in} 
 - \vec p_1^{~in} \cdot \vec p_2^{~in})^2 
- (m_1^{in}m_2^{in})^2  } ,
\label{eobs10}
\end{eqnarray}
and 
\begin{eqnarray}
\tilde A &\equiv& 1 - {E_1 \over E_3} { {\vec p_3 \cdot \vec p_1} \over {p_1^2} } 
\cr 
\tilde B &\equiv& 1 - {E_2 \over E_3} { {\vec p_3 \cdot \vec p_2} \over {p_2^2} } .
\label{eobs11}
\end{eqnarray}

The lab. cross sections follow from the 3N c.m. cross sections
 by the corresponding  Jacobians
\begin{eqnarray}
&& ({d^5\sigma \over {d\Omega_1d\Omega_2dE_1}})^{lab.} = 
 ({d^5\sigma \over {d\Omega_1d\Omega_2dE_1}})^{c.m.} ~ J  \cr  
&& ({d^5\sigma \over {d\Omega_1d\Omega_2dS}})^{lab.} = 
 ({d^5\sigma \over {d\Omega_1d\Omega_2dS}})^{c.m.} ~ 
\vert {  { ({dS\over{dE_1}})^{c.m.}   } 
\over  {  { ({dS\over{dE_1}})^{lab.}   } }}
 \vert ~ J ,
\label{eobs12}
\end{eqnarray}
with 

\begin{eqnarray}
J &=&  { { p_1^{lab.}}  \over { p_1^{c.m.}   } } ~ 
 {  {\vert 
{E_3^{c.m.}\over {p_2^{c.m.}}} - 
{E_2^{c.m.} \over {p_2^{c.m.}}}  {p_3^{c.m.} \over {p_2^{c.m.}}} 
\hat p_3^{c.m.} \cdot \hat p_2^{c.m.} 
\vert} 
 \over { \vert 
{E_3^{lab.}\over {p_2^{lab.}}} - 
{E_2^{lab.} \over {p_2^{lab.}}}  {p_3^{lab.} \over {p_2^{lab.}}} 
\hat p_3^{lab.} \cdot \hat p_2^{lab.} 
\vert     } }\cr
{dS\over{dE_1}} &=& { \sqrt{\tilde A^2 + \tilde B^2} \over {\vert \tilde B \vert} } .
\label{eobs13}
\end{eqnarray}

In Eqs.~(\ref{eobs8}-\ref{eobs13}) $m_i^{in}$ are the rest masses of the incoming particles 
and $E_i^{in}$ ($E_i$) are 
the total energies of the incoming 
 (outgoing) particles. 
The nonrelativistic forms of the cross sections are restored when all total energies are
replaced by the corresponding masses.

\section{Results}

In the following subsections the distribution  over 
the breakup phase-space  of relativistic effects for 
the exclusive breakup cross section will be presented and 
comparison to existing data made. 
The five-fold breakup cross sections 
 can be written 
as 
\begin{eqnarray}
&& {d^5\sigma \over {d\Omega_1d\Omega_2dS}} = (\sum_{m_i m_f} 
\vert < \phi_0 \vert U_0 \vert \phi > {\vert}^2)~\rho_{kin} ,
\label{eobs14}
\end{eqnarray}
with the kinematical factor $\rho_{kin}$ containing the phase-space factor 
and the initial flux.  The differences between the relativistic and 
nonrelativistic cross sections can result from the dynamical part, 
given by the transition probability for breakup,  
 $\sum_{m_i m_f} \vert < \phi_0 \vert U_0 \vert \phi > {\vert}^2$,  
 or/and from the kinematical factor. As a measure of the relativistic effect 
 in a particular complete configuration of the outgoing nucleons  
we take the quantity 
\begin{equation}
\Delta \equiv \Delta(\theta_1, \theta_2, \phi_{12}, S) = 
{ {  ( {d^5\sigma\over {d\Omega_1d\Omega_2dS}})^{rel} 
 - ( {d^5\sigma\over {d\Omega_1d\Omega_2dS}})^{nrel} } \over 
{ ( {d^5\sigma\over {d\Omega_1d\Omega_2dS}})^{nrel}  }  } \times 100\%.
\label{Delta}
\end{equation}

It should be stressed that 
relativistic and nonrelativistic kinematics lead to different S-curves 
in a plane of kinetic energies $E_1^{lab} - E_2^{lab}$. This makes the 
definition of relativistic 
effects for breakup ambiguous and dependent on the procedure which is 
applied to associate points on the relativistic and nonrelativistic 
 S-curves. Those two S-curves can differ significantly in some regions of 
 the breakup phase-space and/or at higher incoming neutron energies. 
 Thus a reasonable projecting procedure is required,   
 especially when the cross sections change drastically along S-curve. 
 In such a case an application of an improper projection method can 
  shift those structures inducing artificially values of $\Delta$. 
 In order   to avoid  such situations 
  we applied the following procedure to associate a point on the 
  nonrelativistic 
S-curve with a given point on the relativistic one.  
 At small values the relativistic and nonrelativistic kinetic energies 
 approach each other and in such a case we projected 
 in $E_1^{lab} - E_2^{lab}$ plane along 
direction perpendicular to the axis $E_i^{lab}$ of the detected 
nucleon with smaller kinetic 
energy. In other  regions  we projected along direction perpendicular to 
the relativistic S-curve. 
 Such a procedure allowed us in most cases to associate properly 
 points on the nonrelativistic S-curve to the points on the relativistic S-curve.

\subsection{Phase-space distribution of relativistic effects}
\label{sub1}

In order to study the distribution of relativistic effects over the breakup 
phase-space we performed the following investigation for the 
d(n,nn)p breakup reaction. At given energy of the incoming neutron 
 we scanned the whole 
breakup phase-space and associated all regions with $\Delta$ values. 
 In order to provide results which could be of interest for future 
experiments we restricted ourselves to regions of phase-space with cross sections 
sufficiently large (for given $\theta_1$, $\theta_2$ and $\phi_{12}$ 
 cross sections smaller than $20\%$ of maximal value have 
been rejected) and kinetic energies of detected neutrons $\ge 5$~MeV. 
 To see how 
relativistic effects depend on energy we carried through such a search 
at two incoming neutron energies $E_n^{lab}=65$~MeV and $200$~MeV. In order 
to locate uniquely the phase-space regions where relativistic effects enhance 
 or diminish the nonrelativistic breakup cross section we 
 show at each energy two sets of 
three two-dimensional plots for positive and negative sign of $\Delta$, 
respectively. The first one is the $\theta_1 - \theta_2$ plane for the two 
angles of the neutron detectors. The second one is the $\theta_1 - \phi_{12}$ 
plane, where $\phi_{12} \equiv \vert \phi_1 - \phi_2 \vert$ is the relative 
azimuthal angle for the two detectors. Finally, the third is the $E_1 - E_2$ 
plane for the correlated lab. kinetic 
 energies of the two detected neutrons. To fill those 
three planes we proceed as follow. The whole phase-space is filled with 
discrete points corresponding to certain grids in $\theta_1$, $\theta_2$, 
 $\phi_1$, $\phi_2$, and $E_1$. For  $\theta_1$ and $\theta_2$ fixed we 
search for the maximal magnitude of $\Delta$ (at given sign of $\Delta$) in 
the three-dimensional subspace spanned by  $\phi_1$, $\phi_2$, and $E_1$. 
Then we combine those maximal $\Delta$ values into four groups and 
associate certain gray tones (colors) to those group values. 
 Next we choose a fixed $\theta_1$ and $\phi_{12}=\vert \phi_2 \vert$ 
(by putting $\phi_1=0$) and search again for the maximal values of 
$\Delta$ in the two-dimensional subspace spanned by $\theta_2$ and $E_1$. 
The same gray tones and groupings are then applied. Finally, in the 
 $E_1 - E_2$ plane we search for the maximal $\Delta$ values in the 
three-dimensional subspace spanned by $\theta_1$, $\theta_2$, $\phi_{12}$ and 
repeat the procedure. 

Results of applying that procedure are shown in the first row of 
Figs.~\ref{fig1}-\ref{fig2} 
 for  $E_n^{lab}=65$~MeV and of Figs.~\ref{fig3}-\ref{fig4} for
$200$~MeV, separately for positive 
(Figs.~\ref{fig1} and \ref{fig3}) and negative 
(Figs.~\ref{fig2} and \ref{fig4}) values of $\Delta$. Distinguishing between 
positive and negative $\Delta$'s allows us to locate regions of breakup 
phase-space where relativity increases or diminishes 
  the nonrelativistic cross 
section. Our numbers are based on the CD~Bonn potential. 
Since we use only gray tones we split the variation of the quantity 
$\Delta$ into four groups. Based on the meaning of the gray tones 
 and using the first row of Figs.~\ref{fig1}-\ref{fig4} one can proceed as 
follows. Choosing a region in the $\theta_1 - \theta_2$ plane with a black 
tone we know that in the $\theta_1 - \phi_{12}$ plane there must exist 
also black region for the same $\theta_1$. This allows to read off 
a certain value of $\phi_{12}$. Then the angular positions of the two 
detectors are fixed, which defines the S-curve in the $E_1 - E_2$ plane. 
Along such an S-curve there must be again a black region, where one can 
read off the corresponding range of energies. Choosing for instance another 
combination of tones, like a black one in the $\theta_1 - \theta_2$ plane, 
 white one in the  $\theta_1 - \phi_{12}$ plane one knows that the S-curve 
in the $E_1 - E_2$ plane lies in the white and maybe gray regions. This should 
explain the use of first row from Figs.~\ref{fig1}-\ref{fig4}. 
 It is seen that relativity can act 
in both directions, increasing or decreasing the 
 nonrelativistic cross section. 
 The  magnitude of relativistic effects increases with energy of the 
 incoming neutron. Whereas at $65$~MeV they approach $\approx 14\%$  
 when relativity increases the nonrelativistic cross section 
  and $\approx 25\%$ in the opposite case, 
 at $200$~MeV the corresponding numbers are $\approx 55\%$ and 
  $\approx 60\%$, respectively. 
  The effects 
are distributed over the entire phase-space. 
 It is seen in Figs.~\ref{fig1}-\ref{fig4} that at both energies 
 (however, at $200$~MeV more clearly) the large  
 relativistic effects have a tendency to localize in phase-space regions  
with small value of the undetected proton energy and when  momenta 
of the two neutrons are coplanar on opposite sides of the beam   
 ($\phi_{12} \approx 180^{\circ}$). 
 Those geometries are around a region 
 of quasi-free-scattering (QFS) condition, where 
the undetected proton is exactly at rest in the lab. system. 
 When relativity increases the nonrelativistic 
cross section ($\Delta > 0$) large relativistic effects   occur at positions 
 of detectors $\theta_1$ and  $\theta_2$ around 
 $\theta_1 + \theta_2 \approx 75^{\circ}$. For  $\Delta < 0$ this region is 
shifted to  
 $\theta_1 + \theta_2 \approx 100^{\circ}$ leading to the following pattern 
 of the nonrelativistic cross 
section variation due to relativity.  
 Starting e.g. at fixed $\theta_2$ from configuration  
where relativity increases the nonrelativistic cross section 
 and increasing $\theta_1$ we are led
to  configurations with decreasing $\Delta$, resulting  finally 
 in geometries where relativity diminishes  the nonrelativistic cross section. 
 In Fig.~\ref{fig5}  we 
show this pattern at $200$~MeV for a number of configurations 
 along S-curve for a fixed angle 
$\theta_2$ and changing $\theta_1$. 
 Of course the same is true when exchanging $\theta_1$ and $\theta_2$.  
This characteristic pattern can be looked for in 
experimental data. 

In order to get insight into the origin of maximal deviations we show in the 
second and the third rows of Figs.~\ref{fig1}-\ref{fig4} the values of 
$\Delta$ calculated from the dynamical and  kinematical factor of 
the cross section, respectively. This is done only for configurations from the first row 
in those figures (geometries with 
 maximal changes of nonrelativistic cross section 
by relativity). It is clearly seen that  the effect 
is predominantly 
 due to the dynamical change of the transition matrix element. For 
localized phase-space regions of the large $\Delta$ values mentioned above 
 the nonrelativistic and relativistic kinematical factors 
are comparable.

\subsection{Comparison to exclusive breakup cross section data}

At higher energies the  
exclusive breakup 
 cross sections have been measured below the pion production threshold 
at $65$~MeV 
\cite{allet1994,allet1996,zejma1997,bodek1998,bodek2001a,kist2003b,kist2005a}, 
 $156$~MeV \cite{br156}, and $200$~MeV \cite{br200}. 
 In Figs.~\ref{fig7}-\ref{fig12} we compare our predictions with the data 
 taken at those energies for some complete geometries. In order to investigate 
 importance of the boost  we show for each geometry the 
nonrelativistic cross section (dotted line) together with  three relativistic 
 cross sections  corresponding to  different treatment of the boost. 
 The first result is 
based on our most extensive approach to the boosted potential 
 (approximation of Eq.~(\ref{eq2apr}): solid line), in the second 
 the boost of the potential from the 2N c.m.  to the 3N c.m. 
is totally neglected (approximation of 
 Eq.~(\ref{ap1}): dashed-dotted line), and in the third 
 the k-dependence of the first order relativistic boost correction 
is omitted (approximation of Eq.~(\ref{ap2}): dashed line). 

For all presented configurations relativistic effects are clearly 
visible. Cases of increasing as well as diminishing the nonrelativistic 
 cross section by relativity are represented. At $65$~MeV (see Fig.\ref{fig7}) 
the smallest effects are for configurations b), c), and d) where both 
nonrelativistic and relativistic predictions provide a satisfactory description 
of data. The largest effects are for two QFS configurations e) and f), where 
the inclusion of relativity reduces the cross section by $\approx 9\%$, and for the
symmetrical space star (SSS) configuration a) where this effect amounts 
to $\approx 7\%$. The SSS geometry corresponds to the situation where for the 
arc-length $S \approx 30$~MeV all three nucleons have equal momenta 
which in the 3N c.m. system lie in the plane perpendicular to the 
beam direction. Despite its smallness 
that relativistic effect leads to a better 
description of data for those three configurations than the nonrelativistic 
result. It seems to explain the small and up to 
now puzzling overestimation of the $65$~MeV SSS cross section 
 data \cite{zejma1997} 
by modern nuclear forces and can account for the experimental width 
of these QFS peaks \cite{allet1994}. For QFS configurations e) and f) 
 the boost effect is significant in the region of QFS peak. Neglecting 
boost decreases the relativistic cross section by  
$\approx 8\%$. However, approximating boost by Eq.(\ref{ap2}) leads 
 practically to the same result as for the full boost.

At $156$~MeV (see Figs.\ref{fig8} and \ref{fig9}) the relativistic effects 
are significantly larger than at $65$~MeV. Here among configurations 
shown there are cases where relativity increases the nonrelativistic 
cross section by up to $\approx 30\%$ and cases with diminished cross 
section by up to  $\approx 50\%$. For those configurations there are regions of 
 kinetic energies $E_1$ where relativity 
 improves the description of data. One can even argue, that 
the pattern described in subsection \ref{sub1} is supported by 
 some of the presented data. Similarly to the situation at $65$~MeV 
importance of the boost depends on configuration and neglecting it changes 
the relativistic cross section up to $\approx 15\%$. 
 It seems that the restriction to the
approximation of Eq.(\ref{ap2}) also here is sufficient for the boost treatment. 

The largest relativistic effects are seen at $200$~MeV 
 (see Figs.\ref{fig10}-\ref{fig12}). 
They can lead to changes of the nonrelativistic cross section increasing 
it even  by up to $\approx 40\%$ or decreasing it by up to 
 $\approx 60\%$,  
 leading predominantly to a better description of data, which  ,however , 
 is not satisfactory in all configurations.  
  Importance of the boost is 
 similar to that  at $156$~MeV. Also here the 
 pattern described in subsection 
\ref{sub1} is discernible. The data of ref.~\cite{br200} shown in 
Figs.\ref{fig10}-\ref{fig12} seem to be shifted in energy $E_1$ with respect 
to both nonrelativistic and relativistic predictions. Independent 
measurement providing cross sections with smaller error bars on the slope 
would be desirable.

\section{Summary and outlook}

We numerically solved  the 3N Faddeev equation for nd scattering 
including relativistic
features such as the relativistic form of the free 
propagator and the change of
the NN potential caused by the boost of the 2N subsystem. The calculations 
 have been performed  
at the  neutron lab. kinetic energies $E_n^{lab} = 65$~MeV, $156$~MeV,   
and $200$~MeV. 
 To construct the 3N momentum space basis we used the relative 
momentum of two nucleons in their 2N c.m. subsystem together with 
 momentum of the spectator nucleon in the 3N c.m. system. 
Such a choice of momenta is adequate for relativistic kinematics 
and allows to generalize the nonrelativistic approach used to solve the  
nonrelativistic 3N
Faddeev  equation  to the relativistic case in a more or less straightforward 
manner. 
That relative momentum in the two-nucleon subsystem is a generalisation of 
the standard
nonrelativistic Jacobi momentum $\vec p$. 
 We neglected the  Wigner
 rotations of the nucleons spins when  boosting the 2N c.m. subsystem to the 
3N c.m. frame.  
  As dynamical input we took
the nonrelativistic NN potential  CD~Bonn  and generated in the 2N c.m. 
system  an exactly on-shell  equivalent relativistic interaction $v$  
 using  the analytical scale transformation of momenta. We checked
 that in our energy range the boost effects for  
 this potential could be sufficiently well incorporated by restricting 
the exact expression 
 to the 
leading order terms in a $q/\omega$ and $v/\omega$  expansion. 
At $65$~MeV and $200$~MeV we performed search for 
magnitudes and signs of relativistic effects on the exclusive nd breakup 
cross sections over the relevant parts of the breakup phase-space. We found, that 
depending on the phase-space region  
relativity can decrease as well as increase the nonrelativistic 
cross section. The magnitude of  effects rises with the incoming neutron 
energy. While at $65$~MeV the effects are rather moderate ($\approx 20\%$) 
at $200$~MeV they can change the nonrelativistic cross section even by a 
factor of $\approx 2$. 
At that energy  relativity leads to 
 a  characteristic pattern of the cross section variation with $\theta_1$ and 
$\theta_2$. Comparison to existing data seems 
to support this finding. At $65$~MeV the inclusion of relativity can explain some 
discrepancies found in the past between theory and data. 

The selectivity of the nd breakup reaction provides opportunity  to study 
different aspects of the 3N dynamics. Since higher energies seem to be more 
favorable to study properties of three-nucleon forces, precise higher 
energy exclusive breakup cross sections should be used as a very valuable tool 
 to test stringently the incorporation of relativity. Especially,
the configurations around the QFS breakup geometry due to their large cross sections and 
insensitivity to the details of nuclear forces seems to be favored 
for this purpose.

\section*{Acknowledgments}
This work has been supported by the  Polish Committee for
Scientific Research under Grant no. 2P03B00825.  
 The numerical
calculations have been performed on the IBM Regatta p690+ of the
NIC in J\"ulich, Germany.


\newpage

\begin{figure}
\caption{(Color online) The regions of the d(n,nn)p breakup phase-space at incoming neutron 
energy $E_n^{lab}=65$~MeV projected onto the $\theta_1 - \theta_2$, 
$\theta_1 - \phi_{12}$, and $E_1 - E_2$ plains, carrying certain 
values of $\Delta$ from Eq.(\ref{Delta}) as indicated in the boxes.  
In the first row the largest positive values of $\Delta$, which measure 
 the largest increasing 
 relativistic effect in the five-fold breakup cross section 
 ${d^5\sigma\over {d\Omega_1d\Omega_2dS}}$, are shown 
 based on the CD~Bonn  potential. 
 The second and the third row shows the values of $\Delta$ obtained only with 
 the  dynamical and kinematical parts of the cross section, respectively, for 
configurations from the first row. The white area belongs either to
phase-space regions not allowed kinematicaly or
to regions rejected by our cuts on the cross sections or energies (see text).
}
\label{fig1}
\end{figure}

\newpage

\begin{figure}
\caption{(Color online) The same as in Fig.~\ref{fig1} but for the largest negative values of 
$\Delta$ at $E_n^{lab}=65$~MeV shown in the first row 
(the largest  decreasing relativistic effect).
}
\label{fig2}
\end{figure}

\newpage

\begin{figure}
\caption{(Color online) The same as in Fig.~\ref{fig1} but for $E_n^{lab}=200$~MeV. 
}
\label{fig3}
\end{figure}
 
\newpage

\begin{figure}
\caption{(Color online) The same as in Fig.~\ref{fig2} but for $E_n^{lab}=200$~MeV.
}
\label{fig4}
\end{figure}

\newpage

\begin{figure}
\includegraphics[scale=0.9]{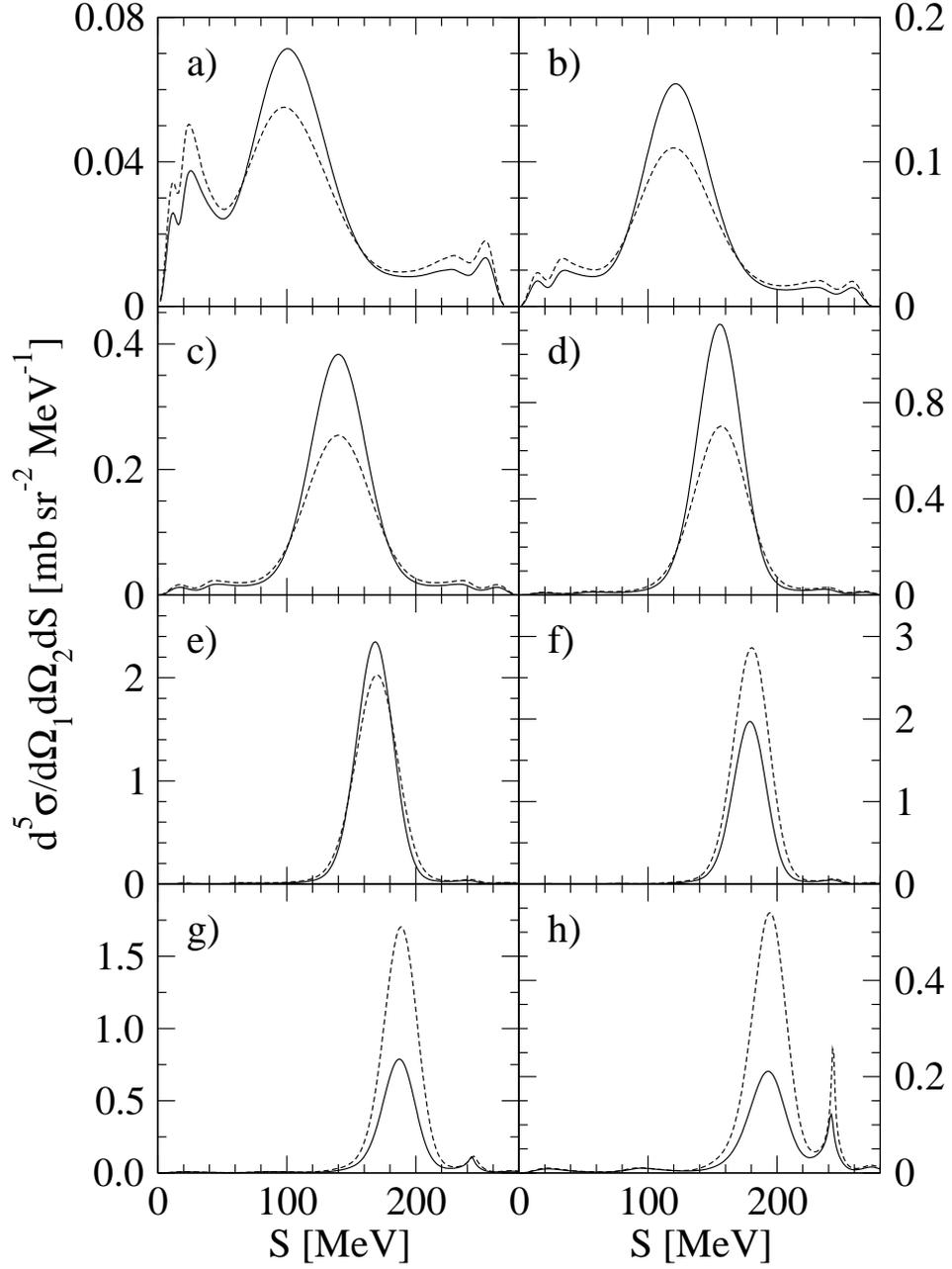}
\caption{The five-fold cross section ${d^5\sigma\over {d\Omega_1d\Omega_2dS}}$ 
for $d(n,n_1n_2)p$ breakup reaction at $E_n^{lab}=200$~MeV at fixed 
 $\theta_2 = 37.5^\circ$, $\phi_{12} = 180^\circ$ and varying  
 $\theta_1$: 
a) $\theta_1 = 27.5^{\circ}$, b) $\theta_1 = 32.5^{\circ}$, 
c) $\theta_1 = 37.5^{\circ}$, d) $\theta_1 = 42.5^{\circ}$, 
e) $\theta_1 = 47.5^{\circ}$, f) $\theta_1 = 52.5^{\circ}$, 
g) $\theta_1 = 57.5^{\circ}$, h) $\theta_1 = 62.5^{\circ}$. 
 The nonrelativistic and relativistic 
cross sections are shown by dashed and solid lines, respectively. 
The calculations are based on the CD~Bonn potential (see text). 
}
\label{fig5}
\end{figure}

\newpage

\begin{figure}
\includegraphics[scale=0.9]{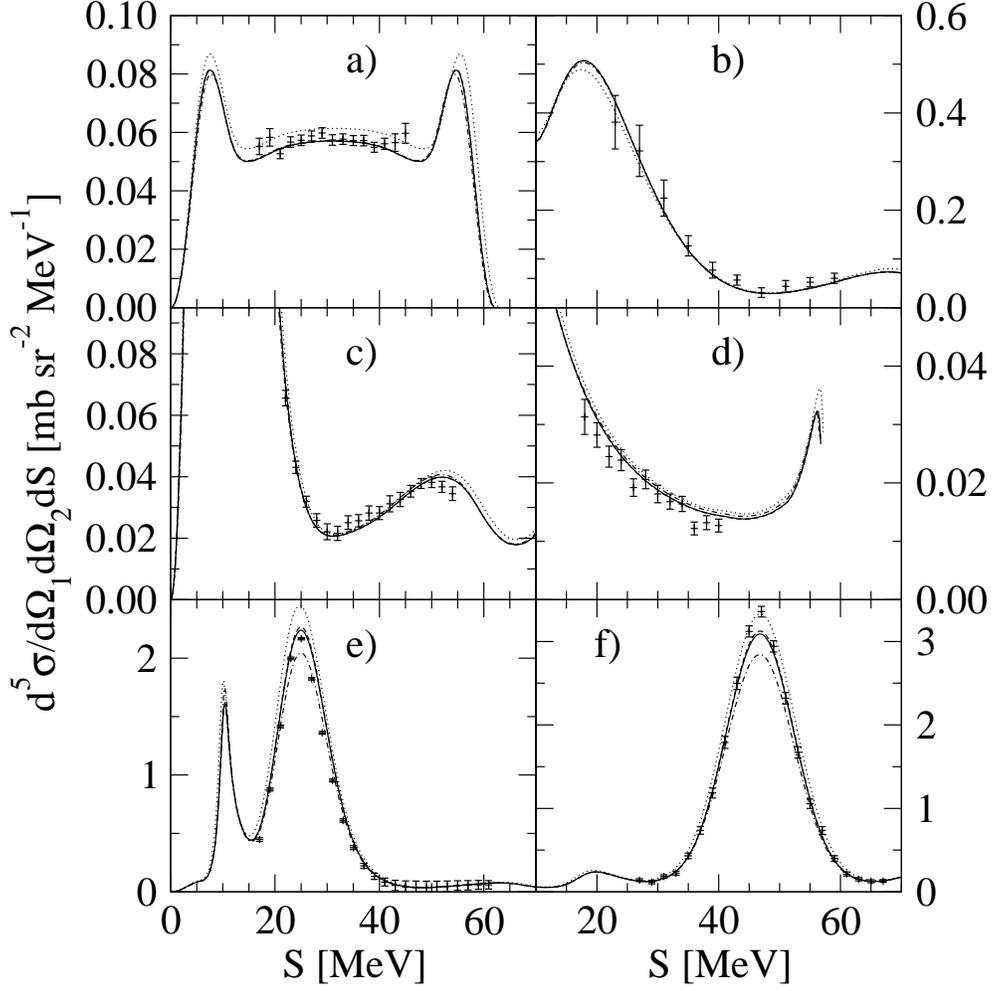}
\caption{The five-fold cross section ${d^5\sigma\over {d\Omega_1d\Omega_2dS}}$ 
for $d(n,n_1n_2)p$ breakup reaction at $E_n^{lab}=65$~MeV for the following 
 positions of outgoing neutron detectors: 
 a) $\theta_1 = \theta_2 = 54^{\circ}$, $\phi_{12} = 120^{\circ}$, 
 b) $\theta_1 = 20^{\circ}$, $\theta_2 = 45^{\circ}$, $\phi_{12} = 180^{\circ}$, 
 c) $\theta_1 = 20^{\circ}$, $\theta_2 = 75.6^{\circ}$, $\phi_{12} = 180^{\circ}$, 
 d) $\theta_1 = 20^{\circ}$, $\theta_2 = 116.2^{\circ}$, $\phi_{12} = 0^{\circ}$, 
 e) $\theta_1 = 30^{\circ}$, $\theta_2 = 59.5^{\circ}$, $\phi_{12} = 180^{\circ}$, 
 f) $\theta_1 = \theta_2 = 44^{\circ}$, $\phi_{12} = 180^{\circ}$. 
 The nonrelativistic and relativistic (full boost) 
cross sections are shown by dotted and solid lines, respectively. 
 Neglecting the boost totally leads to dashed-dotted line. When boost 
 is approximated by 
Eq.(\ref{ap2}) results in dashed line. 
All calculations are based on the CD~Bonn potential (see text). 
The $d(p,pp)n$ data 
are from: ~\cite{zejma1997} a) ; ~\cite{bodek2001a} b), c) and d) ; 
 ~\cite{allet1996} e) and f). 
}
\label{fig7}
\end{figure}

\newpage

\begin{figure}
\includegraphics[scale=0.9]{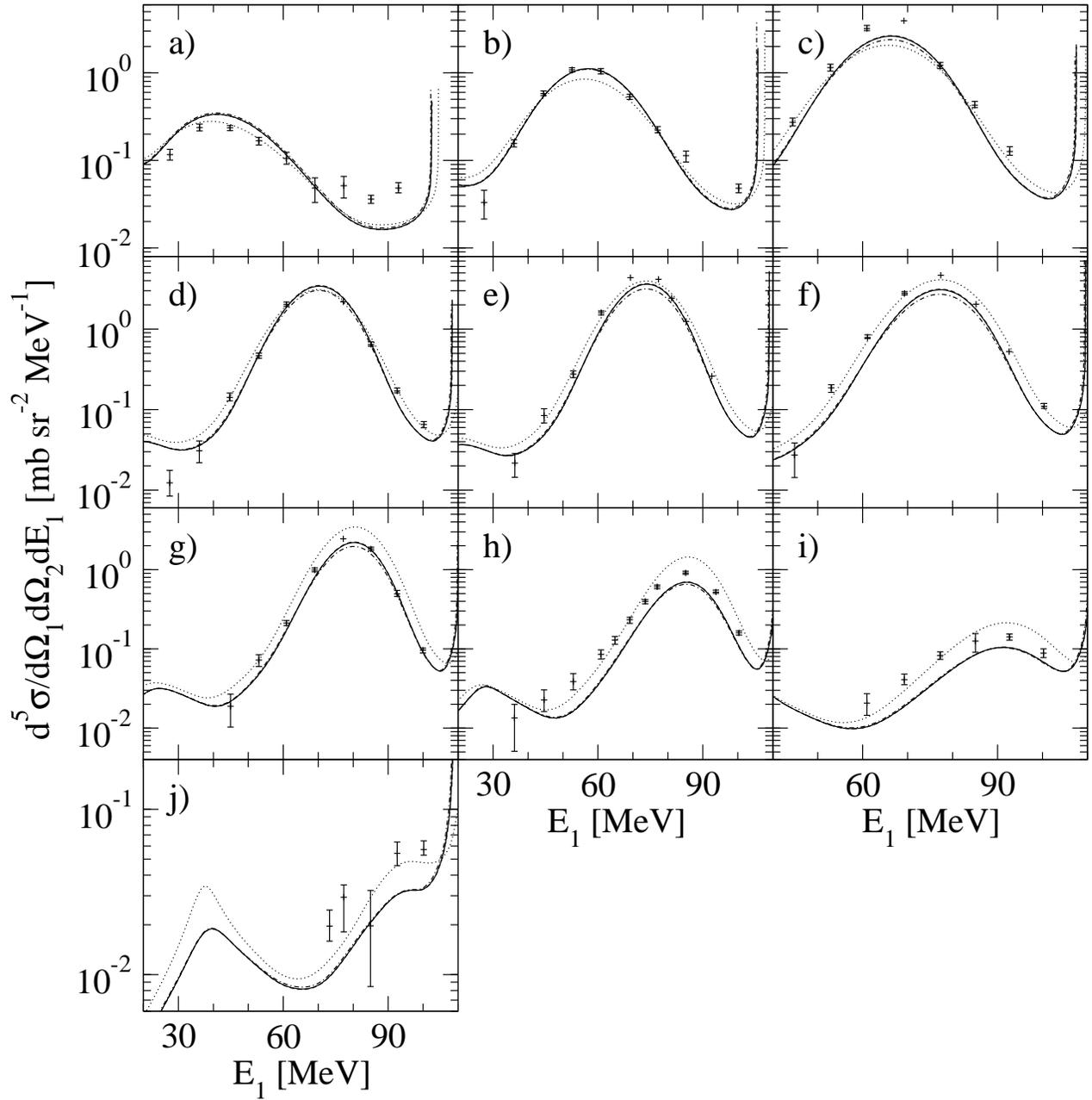}
\caption{The five-fold cross section 
 ${d^5\sigma\over {d\Omega_1d\Omega_2dE_1}}$ 
for $d(n,nn)p$ breakup reaction at $E_n^{lab}=156$~MeV at fixed position 
 $\theta_1 = 45^{\circ}$, $\phi_{12} = 180^{\circ}$, and varying $\theta_2$: 
a) $\theta_2 = 25^{\circ}$, b) $\theta_2 = 32.5^{\circ}$, 
c) $\theta_2 = 37.5^{\circ}$, d) $\theta_2 = 40^{\circ}$, 
e) $\theta_2 = 42.5^{\circ}$, f) $\theta_2 = 45^{\circ}$, 
g) $\theta_2 = 47.5^{\circ}$, h) $\theta_2 = 52.5^{\circ}$, 
i) $\theta_2 = 60^{\circ}$, j) $\theta_2 = 67.5^{\circ}$. 
Curves as in Fig.\ref{fig7}. 
 The $d(p,pp)n$ data 
are from~\cite{br156}. 
}
\label{fig8}
\end{figure}

\newpage

\begin{figure}
\includegraphics[scale=0.9]{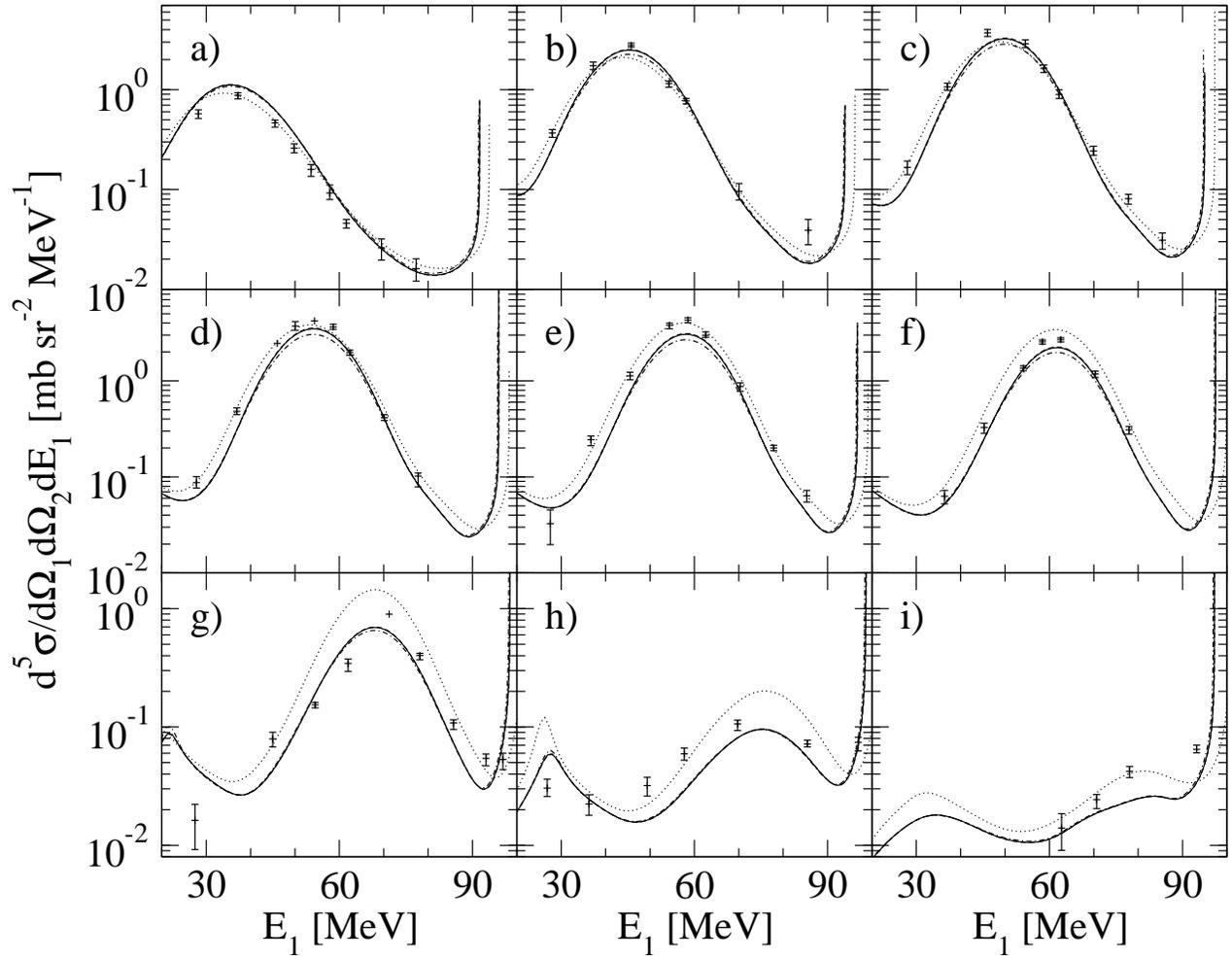}
\caption{The same as in Fig.\ref{fig8} but at fixed position 
 $\theta_1 = 52.5^{\circ}$, $\phi_{12} = 180^{\circ}$, and varying $\theta_2$: 
a) $\theta_2 = 25^{\circ}$, b) $\theta_2 = 30^{\circ}$, 
c) $\theta_2 = 32.5^{\circ}$, d) $\theta_2 = 35^{\circ}$, 
e) $\theta_2 = 37.5^{\circ}$, f) $\theta_2 = 40^{\circ}$, 
g) $\theta_2 = 45^{\circ}$, h) $\theta_2 = 52.5^{\circ}$, 
i) $\theta_2 = 60^{\circ}$. 
}
\label{fig9}
\end{figure}

\newpage

\begin{figure}
\includegraphics[scale=0.9]{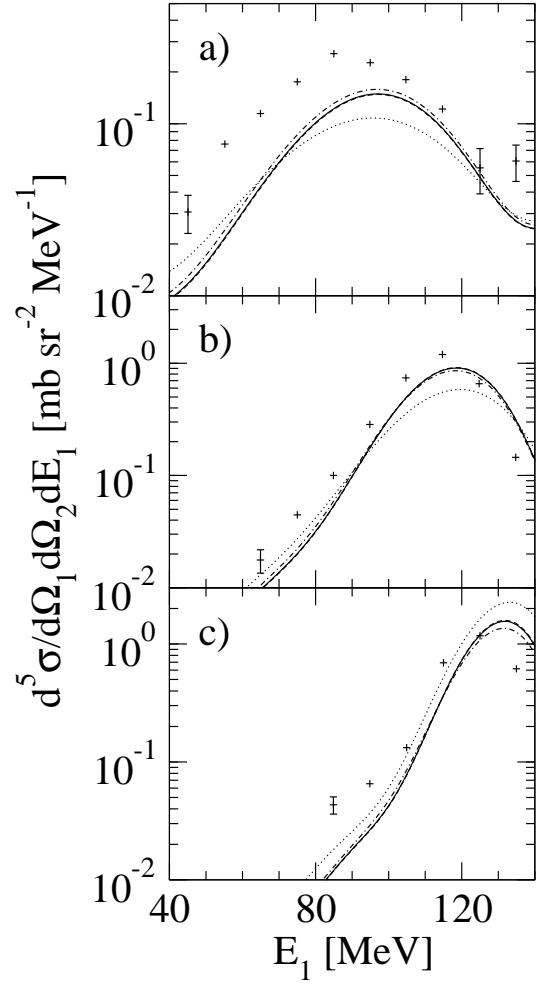}
\caption{The five-fold cross section 
 ${d^5\sigma\over {d\Omega_1d\Omega_2dE_1}}$ 
for $d(n,np)n$ breakup reaction at $E_n^{lab}=200$~MeV  
 at fixed position 
 $\theta_1 = 35^{\circ}$, $\phi_{12} = 180^{\circ}$, and varying $\theta_2$: 
a) $\theta_2 = 35^{\circ}$, b) $\theta_2 = 45^{\circ}$, 
c) $\theta_2 = 55^{\circ}$.
Curves as in Fig.\ref{fig7}. 
 The $d(p,pn)p$ data 
are from~\cite{br200}. 
}
\label{fig10}
\end{figure}

\newpage

\begin{figure}
\includegraphics[scale=0.9]{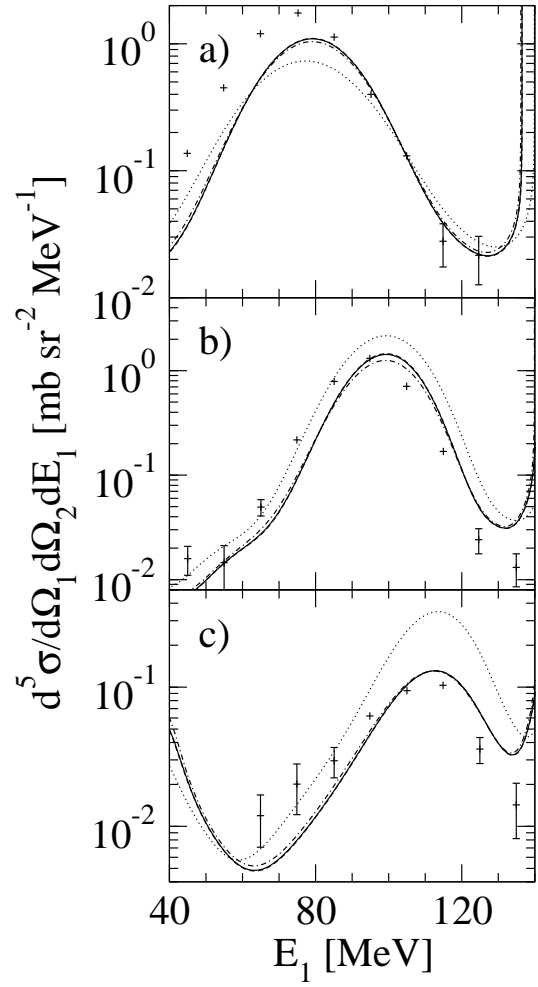}
\caption{The same as in Fig.\ref{fig10} but at fixed position 
 $\theta_1 = 45^{\circ}$, $\phi_{12} = 180^{\circ}$, and varying $\theta_2$: 
a) $\theta_2 = 35^{\circ}$, b) $\theta_2 = 45^{\circ}$, 
c) $\theta_2 = 55^{\circ}$. 
}
\label{fig11}
\end{figure}

\newpage

\begin{figure}
\includegraphics[scale=0.9]{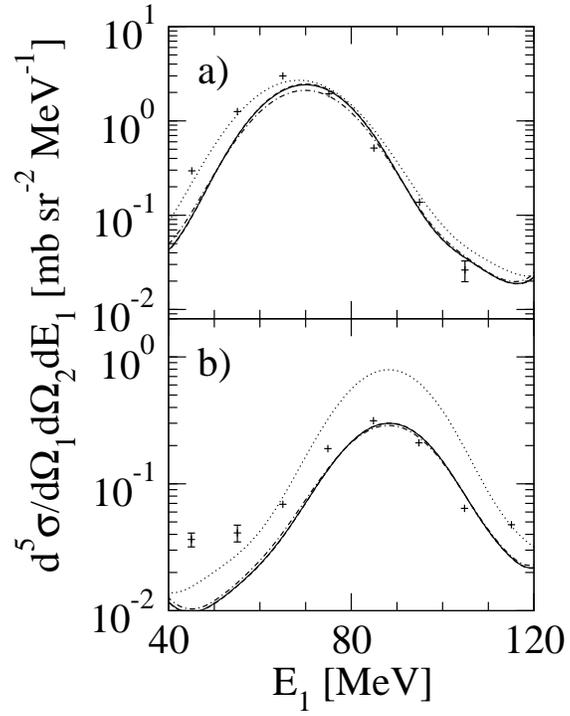}
\caption{The same as in Fig.\ref{fig10} but at fixed position 
 $\theta_1 = 52^{\circ}$, $\phi_{12} = 180^{\circ}$, and varying $\theta_2$: 
a) $\theta_2 = 35^{\circ}$, b) $\theta_2 = 45^{\circ}$. 
}
\label{fig12}
\end{figure}

\end{document}